\documentclass[11pt]{article}
\usepackage{graphicx}
\usepackage{amsmath,amsfonts,amssymb}
\usepackage{color}
\usepackage{bbold}
\def\sloppy{\tolerance=100000\hfuzz=\maxdimen\vfuzz=\maxdimen}
\allowdisplaybreaks

\def \beq  {\begin{equation}}
\def \eeq  {\end{equation}}
\def \beqar {\begin{eqnarray}}
\def \eeqar {\end{eqnarray}}
\def\bsp{\beq\begin{split}}
\def\esp{\end{split}\eeq}
\allowdisplaybreaks
\mathchardef\mhyphen="2D

\def\la {{\langle}}
\def\ra {{\rangle}}

\def\Tr {{\rm Tr}}

\def\del {\partial}
\def\bdel{\bar{\partial}}

\def\A {{\cal A}}

\def\D {{\cal D}}

\def\P {{\cal P}}

\def\half{\textstyle{1\over 2}}

\begin{document}
\def \CMP {{Commun. Math. Phys.}}
\def \PRL {{Phys. Rev. Lett.}}
\def \PL {{Phys. Lett.}}
\def \NPBProc {{Nucl. Phys. B (Proc. Suppl.)}}
\def \NP {{Nucl. Phys.}}
\def \RMP {{Rev. Mod. Phys.}}
\def \JGP {{J. Geom. Phys.}}
\def \CQG {{Class. Quant. Grav.}}
\def \MPL {{Mod. Phys. Lett.}}
\def \IJMP {{ Int. J. Mod. Phys.}}
\def \JHEP {{JHEP}}
\def \PR {{Phys. Rev.}}
\begin{titlepage}
\null\vspace{-62pt} \pagestyle{empty}
\begin{center}
\rightline{CCNY-HEP-16/5}
\rightline{June 2016}
\vspace{1truein} {\Large\bfseries
The Role of the Spin Connection in Quantum Hall Effect:}\\
\vskip .15in
{\Large\bfseries  A Perspective from Geometric Quantization}\\
\vskip .1in
{\Large\bfseries ~}\\
{\large\sc Dimitra Karabali$^a$} and
 {\large\sc V.P. Nair$^b$}\\
\vskip .2in
{\itshape $^a$Department of Physics and Astronomy\\
Lehman College of the CUNY\\
Bronx, NY 10468}\\
\vskip .1in
{\itshape $^b$Physics Department\\
City College of the CUNY\\
New York, NY 10031}\\
\vskip .1in
\begin{tabular}{r l}
E-mail:&{\fontfamily{cmtt}\fontsize{11pt}{15pt}\selectfont dimitra.karabali@lehman.cuny.edu}\\
&{\fontfamily{cmtt}\fontsize{11pt}{15pt}\selectfont vpnair@ccny.cuny.edu}
\end{tabular}

\vspace{.8in}
\centerline{\large\bf Abstract}
\end{center}
\fontfamily{cmr}\fontsize{11pt}{15.6pt}\selectfont
The topological terms of the bulk effective action for the integer quantum Hall effect, capturing the dynamics of gauge and gravitational fluctuations,
reveal a curiosity, namely, 
the Abelian potential for the magnetic field appears in a particular combination with the
Abelian spin connection. This seems to hold for quantum Hall effect on complex projective spaces of arbitrary dimensions. 
An interpretation of this in terms of the algebra of symplectic transformations is given.
This can also be viewed in terms of the metaplectic correction in geometric quantization.

\end{titlepage}

\pagestyle{plain} \setcounter{page}{2}
\setcounter{footnote}{0}
\setcounter{figure}{0}
\renewcommand\thefootnote{\mbox{\arabic{footnote}}}
\fontfamily{cmr}\fontsize{11pt}{15.6pt}\selectfont
\section{Introduction}

There has recently been a lot of research elucidating the effective action
for the quantum Hall effect
on manifolds of different geometries and topologies \cite{general}-\cite{KW2}. This was partly
motivated by the fact that, even though from the experimental
point of view,
we may 
only be interested in spaces of trivial topology, nontrivial geometry and topology can shed light
on various physical quantities such as transport coefficients. 
The mathematical structures involved have also been of interest in their own right.
In two dimensions, the effective action under discussion is best represented as an expansion in powers of the
derivatives of external fields, such as the electromagnetic and gravitational fields.
The leading terms of such a series are topological in character, expressed as a sum of Chern-Simons type terms in the fields. The term involving just the electromagnetic field
and the mixed term involving
both electromagnetic and gravitational fields have been known for a long time \cite{frohlich, WZ}.
The addition of the purely gravitational part and the generalization to include
higher Landau levels revealed an interesting curiosity
\cite{{AG2},{Wiegmann}, {KW2}}.
Apart from the gravitational framing anomaly,
the electromagnetic field and the spin connection of the manifold combine
in a particular way \cite{AG2}.
It is possible to understand the way this combination comes about, both in terms of
isolating the framing anomaly and in terms of the gravitational anomaly due to possible edge modes
in the case of a droplet.
But a more general point of view, based on ideas of geometric quantization, is the subject of this paper.

The quantum Hall effect has also been generalized to higher dimensions \cite{HZ}-\cite{N3} for
a number of different spaces such as the four-sphere \cite{HZ} and complex projective spaces
\cite{KN1}.
Unlike the two-dimensional case, the background gauge fields, the analogue
of the electromagnetic field, can be Abelian or nonabelian.
It is useful to characterize the dynamics of a quantum Hall state
by an effective action.
The part of this effective action which describes the boundary excitations
was obtained in \cite{{KN2}, {poly1}}
as a Wess-Zumino-Witten theory, gauged with respect to the fixed background gauge field.
If fluctuations in the gauge field are possible, there is also nontrivial bulk dynamics.
The
leading terms of the
bulk part  of the effective action in this case are 
topological, being of the Chern-Simons type.
These bulk terms involving the gauge field were given in
 \cite{{Kar},{N3}}, and the general boundary action allowing for fluctuations of the gauge field
was given in \cite{Kar}. Cancellation of anomalies occurs between the bulk and boundary terms.
More recently, we have obtained a general form of the topological terms
of the bulk effective action valid in all dimensions \cite{KN4}, including fluctuations
in the gravitational and gauge fields.
This is done by using the index density in the Dolbeault index theorem as an effective expression for the charge density and then integrating up to obtain the action.
The purely gravitational terms can also be added via the standard descent procedure.
In expanding out the various terms for the complex projective spaces, one again notices the
same curiosity mentioned above: The
Abelian part of the gauge field and the Abelian part of the
spin connection appear in a particular combination.
The recurrence of this combination
 in this generalized context sharpens the need for a
deeper explanation.

It is possible to view the lowest Landau level of a quantum Hall system on a K\"ahler manifold
as the Hilbert space
obtained by the geometric quantization of a symplectic form
which is a suitable multiple of the K\"ahler form.
One of the subtleties of geometric quantization is the appearance
of the metaplectic structure \cite{{geom},{woit},{nair-arm}}.
This arises because we need a quantization procedure which can accommodate
changes of polarization, since physical results should not depend on the polarization 
one uses. This leads to the introduction of half-forms.
The effect of this augmented formalism is that the operator
expressions for certain classical functions get corrections, the so-called metaplectic
correction.
One can also understand this correction in terms of the realization, at the quantum level,
of the algebra of symplectic transformations.
We show that the particular combination of the 
Abelian part of the gauge field and the
spin connection arises in this way. 
These are the main results of this paper.

In the next section, we review the effective action and formulate in more precise terms
the problem we are addressing.
In section 3, we consider the lowest Landau level using geometric quantization and show
the role of the symplectic transformations and
how the metaplectic correction emerges.

\section{ The effective action and the statement of the problem}

We start by recalling some of the essential features of the problem.
We will consider quantum Hall effect on a complex K\"ahler manifold
$K$ of complex dimension $k$ (so that the relevant spacetime is 
${\mathbb R} \times K$). The background gauge fields are valued in the algebra
of $U(k)$ (which is the holonomy group for $K$).
The Abelian part of the background gauge field will be a multiple of the K\"ahler form
$\Omega$ on $K$. (We are interested in the response of the system
to fluctuations of all gauge fields and gravitational fields around the background values).
The standard approach is to set up the Hamiltonian 
for a single particle (corresponding to a field of given spin and charges) and solve the
 Landau problem, construct multiparticle wave functions, etc. However, if we are only interested in the lowest Landau level, the wave functions can be obtained by
 the geometric quantization of a certain symplectic form.
 We will consider these two aspects of the Hall effect here.
 
 The single particle Hamiltonian, apart from any additional potential energy which may be
 needed for confinement of the particles to a droplet,  will be proportional to the 
 Laplace operator on $K$,
 \beq
 H \, \Psi = - {1\over 4 m} (D_{+i} D_{-i} + D_{-i} D_{+i} ) \, \Psi
 \label{met1}
 \eeq
 where $D_{\pm i}$ are (holomorphic/antiholomorphic) derivatives on $K$, suitably covariantized
in terms of their action on $\Psi$.
 The eigenstates of this Hamiltonian fall into distinct Landau levels. The lowest Landau level will obey a
 holomorphicity condition,
 \beq
 D_{-i} \, \Psi_{\rm LLL} = 0
 \label{met2}
 \eeq
 The number of solutions to this condition, and therefore the degeneracy of the lowest Landau level,
  is given by the Dolbeault index theorem \cite{eguchi}.
 Thus, for the case of a completely filled lowest Landau level, where all the available states are occupied by (spinless) electrons, each carrying a unit charge, the index density is identical to the
 charge density, except for terms which can integrate to zero.
 In the case of manifolds which are group cosets, such as for
 $\mathbb{CP}^k = SU(k+1)/U(k)$, the solutions to (\ref{met2}) can be constructed from group
 representation theory \cite{KN2}.
 
While most of the discussion will be of general validity, it is useful to
focus on a specific family of manifolds to see how details work out. We will use
$\mathbb{CP}^k $ for most of what we do.
This manifold has constant Riemannian curvatures valued in the algebra of
$U(k)$, and the background values for the gauge fields are taken to
be proportional to the curvatures. This means also that we can have
an Abelian part for the background gauge field (corresponding to the $U(1)$ part of
$U(k) \sim SU(k) \times U(1)$) and nonabelian gauge fields
valued in $SU(k)$.
The Landau problem of particles in a constant background gauge field is thus
obtained.

Points on the manifold $\mathbb{CP}^k$ can be parametrized by
an element $g$ of $SU(k+1)$, with the identification
$g \sim g\, h, \, h \in U(k) \subset SU(k+1)$, so that
wave functions can be viewed as functions
on $SU(k+1)$ with specified transformation properties under $U(k)$.
Let $t_A$, $A =1, 2, \cdots, k^2+2k$,  denote a basis of hermitian
$(k+1)\times (k+1)$-matrices viewed as the fundamental representation
of the Lie algebra of $SU(k+1)$, with the
normalization
$\Tr \,(t_A t_B )=\half \delta_{AB}$.
The commutation rules of the Lie algebra are of the form
$[t_A, t_B] = i f_{ABC} \, t_C$, with structure constants
$f_{ABC}$.
The generators $t_A$ can be split into
a set of generators for the $SU(k)$ part of
$U(k) \subset SU(k+1)$ (denoted by
$t_a$, $a =1, ~2, \cdots , ~ k^2 -1$) and the
generator for the $U(1)$
direction in $U(k)$ (denoted by 
$t_{k^2+2k}$). The coset generators split into
conjugate sets $t_{\pm i}$, $i= 1, 2, \cdots ,k$.

The matrix elements of $g$ for all the finite-dimensional
representations form a basis for 
functions on the group $SU(k+1)$. These are
the
Wigner $\cal{D}$-functions, which are defined as
\beq
\D^{(J)}_{\mathfrak{l}; \mathfrak{r}}(g) = \la J ,\mathfrak{l}\vert\, g\,\vert J, \mathfrak{r} \ra \label {met3}
\eeq
where $\mathfrak{l}, ~\mathfrak{r} $ stand for two sets of quantum numbers specifying the 
states within the representation.
Further, we can define the
left and right translation operators on $g$ by
\beq
{\hat{L}}_A ~g = T_A ~g, \hskip 1in {\hat{R}}_A~ g = g~T_A
\label{met4}
\eeq
where $T_A$ are the $SU(k+1)$ generators in the representation to which $g$ belongs.

We identify the covariant derivatives  on ${\mathbb{CP}}^k$
in terms of the right translation operators on $g$ as
\beq
D_{\pm i}  = i\,{{\hat R}_{\pm i} \over r}
\label{met5}
\eeq
where $r$ is a parameter with the dimensions of length, defining the scale of the manifold.
The commutator $[ \hat{R}_{+i}, \hat{R}_{-j} ]$ is in the algebra of $U(k)$.
Since this is proportional to the commutator of the derivatives, we can
specify the
constant background fields by the conditions
\beqar
{\hat R}_a ~\Psi^J_{m; \alpha} (g) &=&
(T_a)_{\alpha \beta} \Psi^J_{m; \beta} (g) \label{met6}\\
{\hat R}_{k^2 +2k} ~\Psi^J_{m; \alpha} (g) &=& - {n k\over \sqrt{2 k
(k+1)}}~\Psi^J_{m; \alpha} (g) \label{met7}
\eeqar
where $m=1,\cdots, {\rm dim}J$ gives the degeneracy of the Landau level.
The wave functions $\Psi^J_{m; \alpha}$ transform on the right
as a representation ${\tilde J}$
of $SU(k)$, 
$(T_a)_{\alpha \beta}$ being the representation matrices.
Likewise, (\ref{met7}) shows that $\Psi^J_{m; \alpha}$ carry a particular charge
for $U(1) \subset U(k)$;
$n$ is the strength
of the Abelian part of the background gauge field. (The corresponding field strength is $n \Omega$, where $\Omega$ is the K\"ahler form and $n$ is an integer by the Dirac quantization condition.)
$\alpha ,\beta$ label states within the $SU(k)$ representation ${\tilde J}$
(which is itself
contained in the representation $J$ of $SU(k+1)$). 
The index $\alpha$ carried by the
wave functions $\Psi^J_{m; \alpha} (g)$
 is basically the gauge index. The wave functions are sections
of a $U(k)$ bundle on ${\mathbb{CP}}^k$.
By virtue of (\ref{met5}, \ref{met6}, \ref{met7}), we can write
\beq
H \, \Psi 
= {1\over 2 m r^2} \left[ \hat{R}_{+i} \hat{R}_{-i} +  {i \over 2} f_{-i, +i, a} \, T_a + 
{i \over 2} f_{-i, +i, k^2+2k} \, \left( - {n k \over \sqrt{2 k (k+1)}} \right) \right] \Psi
\label{met8}
\eeq
The Hamiltonian $H$ is proportional to 
$\sum_{i} {\hat R}_{+i} {\hat R}_{-i}$, apart from additive constants.
The lowest Landau level evidently satisfies
\beq
\hat{R}_{-i} \, \Psi = 0
\label{met9}
\eeq
This is the holomorphicity condition (\ref{met2}) in
terms of the group translation operators.
Writing $\Psi^J_{m; \alpha} (g) \sim \la J, m\vert g \vert J, \alpha , w\ra$, the
conditions (\ref{met9}, \ref{met6}, \ref{met7}) become
\beqar
\hat{R}_{-i} \, \vert J, \, \alpha, w \ra &=& 0\label{met10}\\
\hat{R}_a \, \vert J, \, \alpha, w \ra = (T_a)_{\alpha \beta} \, \vert J, \, \beta, w \ra,
&\hskip .1in&
\hat{R}_{k^2 + 2k}  \, \vert J, \, \alpha, w \ra = - {n \, k \over \sqrt{2 k (k+1)}}
\, \vert J, \, \alpha, w \ra
\label{met11}
\eeqar
According to (\ref{met10}), for the lowest Landau level, the
 state $\vert J, \, \alpha, w \ra$ must be a lowest weight state in the 
representation $J$, with weight $w= {-n \, k \over \sqrt{2 k (k+1)}}$, specified by
(\ref{met11}). The representation $J$ is completely fixed by
(\ref{met10}), (\ref{met11}).

We now recapitulate the essential features of the effective action from
\cite{KN4}.
In that paper, we argued that higher Landau levels for spinless electrons, say the $s$-th level,
could be viewed for the purpose of the effective action, as the lowest Landau level for higher spin fields. For the case of $\mathbb{CP}^k$, these higher spin fields couple to the constant background field of the form
\beq
\bar{\cal F} =-i \bigl( n\,\Omega\, \mathbf{1} + s \bar{R}^0 \mathbf{1} + \bar{R}^a T_a \bigr) = \bar{F} + \bar{\cal{R}}_s
\label{met12}
\eeq
where $\bar{R}^0,~\bar{R}^a$ are the curvature components for $\mathbb{CP}^k$
corresponding to the
$U(1)$ and $SU(k)$ subgroups of the holonomy group $SU(k+1)$
and $T_a,~\mathbf{1}$ are $U(k)$ matrices in the appropriate spin representation, $s$ being the
$U(1)$ spin.
We will also include fluctuations around these background values in what follows.
The strategy in \cite{KN4} was to consider the number of solutions to
the holomorphicity condition (\ref{met2}) as given by the Dolbeault index theorem
\cite{eguchi},
\beq
{\rm Index}(\bdel_V) = \int_K {\rm td}(T_cK) \wedge {\rm ch} (S \otimes V) 
\label{met13}
\eeq
where ${\rm ch}$ denotes the Chern character given by
\beq
{\rm ch} (S \otimes V) = \Tr \left(  e^{i ({\cal R}_s +F) /2 \pi} \right) 
= {\rm ch} (S) \wedge {\rm ch} (V)
\label{met14}
\eeq
In this equation, ${\cal R}_s$ is the curvature in the representation 
appropriate to the chosen spin 
and $F$ is in the representation for the (gauge) charge rotations
of the particles under consideration.
Further, in (\ref{met13}),  ${\rm td}(T_cK)$ denotes the Todd class for the complex tangent bundle of 
the phase space, given explicitly by traces of products of curvatures.
Explicit formulae are given in many places, including \cite{eguchi} and \cite{KN4}.
Taking the index density as the charge density we can derive the effective action
for a completely filled lowest Landau level by ``integrating" the index density with respect to the time-component of the Abelian gauge field $A_0$ and making the result covariant \cite{KN4}. The effective action is then given by 
\beq
S^{(s)}_{ 2k+1} =
 \int \Bigl[ {\rm td}(T_c K) \wedge \sum_p  (CS)_{2 p+1} (\omega_s + A)\Bigr]_{2 k+1}
+ 2 \pi\int \Omega^{\rm grav}_{2k+1}
+ {\tilde S}
\label{met15}
\eeq
Here $\omega_s$ is the spin connection corresponding to ${\cal R}_s$ and 
$A$ is the connection for the gauge field $F$. $\Omega^{\rm grav}_{2k+1}$
is defined by
\beq
\left[ {\rm td}(T_cK) \wedge {\rm ch} (S)\right]_{2 k +2}
=  d\, \Omega_{2 k+1}^{\rm grav} + {1\over 2 \pi} \, d\, \Bigl[ {\rm td}(T_c K) \wedge \sum_p  (CS)_{2 p+1} (\omega_s )\Bigr]_{2k+1}
\label{met16}
\eeq
Thus $d\,\Omega_{2 k+1}^{\rm grav} $ gives the $(2k+2)$-form in ${\rm td} (T_cK)$.
Further, we note that the Chern-Simons term is related to the curvatures by
\beq
{1\over 2 \pi} d (CS)_{2 p +1} (A) = {1 \over (p+1)!} \Tr \left(
{ i F \over 2 \pi}\right)^{p +1}
\label{met17}
\eeq
Also, ${\tilde S}$ in (\ref{met15}) refers to nontopological terms including those
due to the fact that the charge density could differ from the index density as given by 
the integrand in (\ref{met13}) by terms which are total derivatives integrating to zero.
These terms are expected to be of higher order in a derivative expansion for the
external fields.

Various special cases of this action have been discussed in \cite{KN4}. 
For the present discussion, we will
consider $\mathbb{CP}^k$, $k = 1, 2, 3$.
For simplicity, we will consider
only the spinless case so that $s = 0$ (i.e. only the lowest Landau level)
with the background gauge fields being
purely Abelian (valued in $\underline{U(1)}$). This will suffice to illustrate the
main point.
For the $2+1$ dimensional case, the action becomes
\beq
S_{3d} =
 {i^2 \over {4\pi}} \int \Biggl\{\Bigl(A + {1 \over 2}\,\omega\Bigr) \, d\Bigl(A + {1 \over 2}\,\omega\Bigr) - {1 \over 12} \omega \,d\omega \Biggr\}
\label{met18}
\eeq
(We may note that this result agrees with \cite{AG2}-\cite{KW2} as well.)
In $4+1$ dimensions, we have
\beq
S_{5d} = {i^3\over {(2\pi)^2}} \int \Biggl\{ { 1 \over 3!} \Bigl(A+ \omega^0\Bigr) \Bigl[d\Bigl(A+\omega^0\Bigr)\Bigr]^2 
-{ 1 \over 12} \Bigl(A+ \omega^0\Bigr) \Biggl[  (d\omega^0)^2  +{1\over 2} \Tr ( {\tilde R} \wedge {\tilde R}) \Biggr] \Biggr\}
\label{met19}
\eeq
where ${\tilde R}$ is the $SU(2)$ part of the
curvature and $\omega^0$ is the $U(1)$ part of the spin connection.
In $(6+1)$ dimensions, the effective action is
\beqar
S_{7d} &=& { 1 \over (2\pi)^3} \int \Biggl\{ { 1 \over 4!} \left(A + {3 \over 2} \omega^0\right) \left[ d\left( A + {3 \over 2} \omega^0\right)\right]^3\nonumber\\
&&\hskip .7in - {1 \over 16} \left(A + {3 \over 2} \omega^0\right)d\left(A + {3 \over 2} \omega^0\right) \left[ (d\omega^0)^2 + {1 \over 3} \Tr (\tilde{R} \wedge \tilde{R}) \right] \nonumber \\
&&\hskip .7in +{ 1 \over 1920} \omega^0d\omega^0 \left[ 17 (d\omega^0)^2 + 14 \Tr (\tilde{R} \wedge \tilde{R}) \right] + {1 \over 720} \omega^0 \Tr (\tilde{R} \wedge \tilde{R}\wedge \tilde{R}) \Biggr\} 
\nonumber\\
&&+ {1 \over 120} \int (CS)_7 (\tilde{\omega})
\label{met20}
\eeqar
where ${\tilde R}$ is now the $SU(3)$ curvature and $\tilde{\omega}$ the corresponding connection.

The fields $A$, $\omega^0$, ${\tilde \omega}$ in (\ref{met18}-\ref{met20})
include fluctuations around the background values pertinent to $\mathbb{CP}^k$.
Notice that the gauge field appears in the combination
$A + {k \over 2} \omega^0$.
Further, even if we set the combination $A + {k \over 2} \omega^0$ to zero, there are purely gravitational terms in (\ref{met18}-\ref{met20}) for $d = 2 +1 $ and $6+1$, not for
$d = 4+1$.
It may be possible to understand these left-over purely gravitational terms
in terms of 
the gravitational anomaly due to boundary excitations.
Here we are still considering a closed manifold with no boundary, but if we
think of enlarging the context by considering a droplet of fermions
of finite size,
excitations on the edge or boundary of the droplet are possible.
These edge modes would be described by a chiral theory
in $(2 k -1, 1)$ dimensions, and such a theory can have
a gravitational anomaly only if $k$ is an odd integer
\cite{ag-witten}. The cancellation of the anomaly between the boundary
and the bulk would necessitate purely gravitational bulk terms.
Once such terms are identified and isolated, it should be possible to see
why the remainder of the action involves the combination $A+ {k \over 2} \omega^0$.
Analysis from this point of view, in two dimensions, has been carried out in 
\cite{{AG2}, {Wiegmann}}.

But we can ask: Is there an independent way of
seeing why
the combination
$A + {k \over 2} \omega^0$ is natural?
This is the question we seek to address in this paper.
If such an argument works out, we may be able to
utilize this to shed some light on the nature of the edge modes,
if we propose to consider a droplet.
\section{The perspective of geometric quantization}

As mentioned in the last section, the
second way to think about this problem is to focus on the lowest Landau level
and obtain the wave functions via geometric quantization \cite{geom}.
We will be interested in the case of
$K = \mathbb{CP}^k$,
with the background gauge field being entirely Abelian; i.e.,
we have a trivial representation for $SU(k)$ in (\ref{met11}). 
For the geometric quantization of $\mathbb{CP}^k$,
we can consider the
symplectic form $n \, \Omega$, where 
$\Omega$ is the K\"ahler form.
Upon quantization, this
leads to the lowest Landau level as given by (\ref{met10}, \ref{met11}). 
The holomorphicity
condition (\ref{met10}) becomes the Bargmann (or K\"ahler) polarization condition on the
wave functions.
An alternative approach is 
to consider the flat space $\mathbb{C}^{k+1}$, use the obvious symplectic form
 on this space, carry out the quantization and then reduce via a constraint
 to obtain results relevant to the projective space.
 
 There are then slightly different ways to argue for the emergence of the combination
 $A + (k/2) \omega^0$. One way is to start with $\mathbb{C}^{k+1}$, quantize and then require the implementation of a set of symplectic transformations. This can be done via the operators 
 realizing the
 algebra of the symplectic transformations.The closure of the algebra will naturally lead to
 $U(k+1)\sim SU(k+1) \times U(1)$ transformations with a modified operator for the 
 $U(1)$ part. This will ultimately  lead to the combination  $A + (k/2) \omega^0$.
 We can then argue that the generators of $U(k+1)$ descend to the
 $\mathbb{CP}^{k+1}$ space of interest.
 Another approach would be to consider $\mathbb{C}^{k+1}$ again, and obtain the correction
 to the generator of the $U(1)$ from ``half-forms".
 Again, one can argue that this descends to $\mathbb{CP}^k$.
 We will consider these two related ways in turn.
 A third approach would
 be to directly start with $\mathbb{CP}^k$ classically and then quantize using
 ``half-forms" and obtain the corrected operators of interest. We will  not pursue this here, but it remains an interesting question. 

We start with the following symplectic two-form on $\mathbb{C}^{k+1}$,
\beq
M = i \, d Z_\alpha \wedge d{\bar Z}_\alpha .
\label{met28}
\eeq
We can then impose the constraint
\beq
{\bar Z}_\alpha Z_\alpha - c = 0
\label{met29}
\eeq
for some constant $c$.
The symplectic reduction of $\mathbb{C}^{k+1}$
by this constraint leads to $\mathbb{CP}^k$.
In other words, (\ref{met29}) is to be viewed as a first class constraint in the sense of
Dirac's theory of constraints.
The condition
(\ref{met29}) reduces the space $\mathbb{C}^{k+1}$
 to the sphere $S^{2 k +1}$ and a gauge-fixing constraint conjugate to
(\ref{met29}) eliminates an overall phase for the $Z$'s, giving
$\mathbb{CP}^k$ as $S^{2 k+1}/ S^1$.

The quantization of (\ref{met28}) in the holomorphic polarization leads to the usual
coherent state wave functions
\beq
\Psi = \exp\left( - {\half} {{\bar Z}\cdot Z}\right) ~ h(Z)
\label{met30}
\eeq
where $h(Z)$ is holomorphic.
The operators corresponding to $Z_\alpha$, ${\bar Z}_\alpha$ are 
$a^\dagger_\alpha$, $a_\alpha$ respectively, with $[ a_\alpha , a^\dagger_\beta]
= \delta_{\alpha\beta}$. The coherent states are of the form
\beq
\vert {\bar Z}\ra = \exp\left( - {\half} {{\bar Z}\cdot Z}\right) ~ e^{ {\bar Z} \cdot a^\dagger}
\vert 0\ra
\label{met31}
\eeq
where $\vert 0\ra$ is the Fock vacuum, $a_\alpha \vert 0\ra = 0$.
The quantum version of
constraint (\ref{met29}) is of the form
$a^\dagger\cdot a - c'$, for some value $c'$. We consider the reduction of the 
Hilbert space for (\ref{met28}) by this constraint, choosing a particular value
$c' =n$. This means that the states should now obey
\beq
( a^\dagger \cdot a - n) \, \vert n \ra = 0,
\label{met31a}
\eeq
so that
$\vert n\ra$ is given by
$a^\dagger_{\alpha_1} a^\dagger_{\alpha_2} \cdots a^\dagger_{\alpha_n} \vert 0\ra$.
The wave functions corresponding to this are, up to normalization,
\beq
\Psi \sim \la Z \vert \, a^\dagger_{\alpha_1} a^\dagger_{\alpha_2} \cdots a^\dagger_{\alpha_n} \vert 0\ra
\sim Z_{\alpha_1} Z_{\alpha_2} \cdots Z_{\alpha_n} ~e^{- {\half} {\bar Z}\cdot Z}
\label{met32}
\eeq
If we relate $Z_\alpha$ to an $SU(k+1)$ element $g_{\alpha \, k+1}$
via $Z_\alpha = \lambda \, g_{\alpha \, k+1}$, then these wave functions are seen to
be proportional to the Wigner functions
$\la J, \mathfrak{l} \vert g \vert J, 0, w\ra$; the functions
$\Psi \sim \la J, \mathfrak{l} \vert g \vert J, 0, w\ra$ satisfy
(\ref{met6}) and (\ref{met7}) with ${\hat R}_a \, \Psi = 0$.

Let us now start again with (\ref{met28}) before the imposition of
the constraint (\ref{met29}).
Rather than using the complex coordinates $Z, \, {\bar Z}$, let us consider
using real coordinates $p_\alpha, \, q_\alpha$, with
\beq
Z_\alpha = {1\over \sqrt{2}}\, ( p _\alpha + i q_\alpha),
\hskip .2in
{\bar Z}_\alpha = {1\over \sqrt{2}}\, ( p _\alpha - i q_\alpha)
\label{met32a}
\eeq
This is equivalent to viewing $\mathbb{C}^{k+1}$ as $\mathbb{R}^{2 k +2}$; the
two-form $M$ is now $M = dp_\alpha \wedge dq_\alpha$.
One could also consider new complex combinations, say, $\xi_\alpha,  \, {\bar \xi}_\alpha$
 of $p_\alpha,\, q_\alpha$, other than
the ones in (\ref{met32a}), and consider the holomorphic quantization
of $M$, holomorphicity being defined by the new choice. For example,
if we choose
\beqar
\xi_\alpha &=& {1\over \sqrt{2}}\, \left[ p_\alpha + G_{\alpha\beta}\, p_\beta
+ H_{\alpha\beta} \, p_\beta + i \left( q_\alpha + G_{\alpha\beta}\, q_\beta
- H_{\alpha\beta}\, q_\beta\right) \right]\nonumber\\
{\bar \xi}_\alpha &=& {1\over \sqrt{2}}\, \left[ p_\alpha + G^*_{\alpha\beta}\, p_\beta
+ H^*_{\alpha\beta} \, p_\beta - i \left( q_\alpha + G^*_{\alpha\beta}\, q_\beta
- H^*_{\alpha\beta}\, q_\beta\right) \right]
\label{met32b}
\eeqar
where $G_{\alpha\beta}$ is antihermitian, $G^*_{\alpha\beta} = - G_{\beta \alpha}$,
 and $H_{\alpha\beta}$ is symmetric,
 it is easily verified that
 \beq
 M = dp_\alpha \wedge dq_\alpha
 = i \, d\xi_\alpha \wedge d{\bar \xi}_\alpha
 \label{met32c}
 \eeq
to linear order in $G$, $H$. If we quantize using coherent states
defined by the $\xi,\, {\bar \xi}$ or by
the original $Z,\, {\bar Z}$, the quantum theory should be the same, since they both
correspond to the same $M = dp_\alpha \wedge dq_\alpha$.
This means that we should be able to implement the change
from $Z,\, {\bar Z}$ to $\xi,\, {\bar \xi}$ by a unitary transformation
in the quantum theory.

To see how this works out, we first
 write $\xi_\alpha$, ${\bar \xi}_\alpha$ directly in terms of
the $Z_\alpha$, ${\bar Z}_\alpha$ as
\beq
\left( \begin{matrix}
\xi_\alpha\\ {\bar \xi}_\alpha \\
\end{matrix}\right) = \left\{ 
\left[ \begin{matrix} \delta_{\alpha\beta} &0 \\ 0&\delta_{\alpha\beta}\\
\end{matrix}\right] + 
\left[ \begin{matrix} G_{\alpha\beta} & H_{\alpha\beta} \\ 
H^*_{\alpha\beta}&G^*_{\alpha\beta}\\
\end{matrix}\right]  \right\} \, 
\left( \begin{matrix}
Z_\beta\\ {\bar Z}_\beta \\
\end{matrix}\right)
\label{met33}
\eeq
This is the infinitesimal transformation, since we only kept
$G_{\alpha\beta}$, $H_{\alpha\beta}$ to linear order in verifying
(\ref{met32c}). But finite transformations can be constructed
by a sequence of infinitesimal transformations and they too preserve
(\ref{met32c}).
The finite transformations corresponding to (\ref{met33}) form the symplectic
group $Sp (k+1, \mathbb{R})$.
The classical generating function for the $G$ and $H$-type transformations are
\beq
G =  i \,G_{\alpha \beta} Z_\beta {\bar Z}_\alpha , \hskip .2in
H = {i\over 2} H_{\alpha \beta} {\bar Z}_\alpha
{\bar Z}_\beta ,
\label{met33a}
\eeq
respectively.
The quantum version of these are the operators
\beqar
{\hat G} &=& i \, G_{\alpha \beta} \, a^\dagger_\beta\, a_\alpha ~+~{\rm ordering ~ambiguities}\nonumber\\
{\hat H} &=& {i \over 2} H_{\alpha \beta} \, a_\alpha \, a_\beta
\label{met34}
\eeqar
There are ordering ambiguities for ${\hat G}$, affecting the terms
with $\alpha = \beta$.
Classically, the Poisson bracket of ${i\over 2} H_{\alpha \beta} {\bar Z}_\alpha
{\bar Z}_\beta$ and its conjugate gives the generators of the $G$-type transformation,
the full algebra being the Lie algebra of $Sp (k+1, \mathbb{R})$.
Quantum mechanically, commuting the generator of the $H$-type transformation and its 
conjugate we find
\beqar
[ a_\alpha a_\beta , a^\dagger_\gamma a^\dagger_\delta ]
&=& \left( \delta_{\alpha\gamma} J_{\delta \beta} +  \delta_{\alpha\delta} J_{\gamma \beta} 
+  \delta_{\beta\gamma} J_{\delta \alpha}  +  \delta_{\beta\delta} J_{\gamma\alpha} 
\right)
+ {2 \over k+1} \left( \delta_{\alpha\gamma} \delta_{\beta \delta}
+  \delta_{\alpha\delta} \delta_{\beta \gamma} \right)~Q\nonumber\\
J_{\alpha \beta} &=& a^\dagger_\alpha a_\beta - {\delta_{\alpha\beta} \over k+1}
a^\dagger \cdot a \nonumber\\
Q&=& a^\dagger \cdot a + {\half} ({k +1})
\label{met35}
\eeqar
$J_{\alpha\beta}$ are the generators of $SU(k+1)$ and $Q$ generates a $U(1)$ transformation.

We now want to consider the reduction to $\mathbb{CP}^k$.
The key point is that while the generator ${\hat H}$ and its conjugate
do not commute with the constraint
$a^\dagger\cdot a - n$, 
$J_{\alpha\beta}$ does commute with it. Thus we expect the action of
$J_{\alpha\beta}$ to descend to the case
of $\mathbb{CP}^k$.
In fact, 
the $SU(k+1)$ transformations generated by $J_{\alpha\beta}$
 are the isometries of the reduced
space. Similarly $Q$ commutes with the constraint
$a^\dagger\cdot a - n$ and we should expect its action to descend to
$\mathbb{CP}^k$ as well. The key point is that the
Lie algebra of $Sp(k+1, \mathbb{R})$ at the level of
$\mathbb{C}^{k+1}$
chooses a certain operator ordering,
giving the unambiguous quantum expressions for the generators, before we consider their descent
to $\mathbb{CP}^k$.
(The relevance of the $Sp(k+1, \mathbb{R})$ in quantizing
(\ref{met28}) is discussed in \cite{woit}. Our main point is that since
$J_{\alpha\beta}$, $Q$ commute with the constraint
(\ref{met29}), we can easily adapt that discussion
to the case of $\mathbb{CP}^k$.)

From the commutator of $Q$ with $a_\alpha$ and $a^\dagger_\alpha$, we see that
it generates the phase transformation,
\beq
e^{i Q \theta} \, a_\alpha \, e^{-i Q \theta} = e^{-i \theta}~ a_\alpha,
\hskip .2in
e^{i Q \theta} \, a^\dagger_\alpha \, e^{-i Q \theta} = e^{i \theta}~ a^\dagger_\alpha
\label{met36}
\eeq
Classically, this is the transformation
 $Z_\alpha \rightarrow e^{i\theta} Z_\alpha$,
${\bar Z}_\alpha \rightarrow e^{-i\theta} {\bar Z}_\alpha$.
The product
$Z_{\alpha_1} Z_{\alpha_2} \cdots Z_{\alpha_n}$ gets an overall phase
$e^{i \theta n}$. However, notice that $Q$ has a value
$n + {\half} (k+1)$ for
 the state 
$a^\dagger_{\alpha_1} a^\dagger_{\alpha_2} \cdots a^\dagger_{\alpha_n} \vert 0\ra$.
Thus there is an additional ``zero-point" value for $Q$.
This is the ``correction" we are after.

To complete this part of the story, we now show that
this extra ``zero-point" charge couples to the spin connection
when gravitational fluctuations are introduced.
The identification $Z_\alpha \sim  \,g_{\alpha\, k+1}$ shows that
the phase transformation $Z_\alpha \rightarrow e^{i \theta} Z_\alpha$
is equivalent to a  right transformation of $g$ by an element of
$U(1) \subset U(k)$. This $U(1)$ is the transformation generated by
$\hat{R}_{k^2 + 2k}$.
In the description of $\mathbb{CP}^k$ as $SU(k+1)/U(k)$,
with coordinates given by $g_{\alpha \, k+1} \in SU(k+1)$, the
right action by
$\hat{R}_{k^2+2k}$, $\hat{R}_a$ generate the isometries.
Thus the $U(1)$ under discussion does correspond to the
$U(1)$ part of the
isometry group; hence  its gauging is indeed 
done by the $U(1)$ spin connection.
Further, the
background magnetic field, chosen to be proportional
to the spin connection and specified by (\ref{met11}) leads to the
monomial $Z_{\alpha_1} Z_{\alpha_2} \cdots Z_{\alpha_n}$.
Since we do have the extra $Q$ charge even in the absence of a magnetic field, we must
interpret this extra charge ${\half} (k+1)$ as the coupling constant for the spin connection. Thus we expect the combination
$\alpha\, (n+ {\half}(k+1)) \omega^0$ where $\alpha$ takes care of any overall
normalization for the fields.
Actually our chosen normalization for the spin connection was such that
$d \omega^0 = {k+1\over k} \Omega$, while
the gauge field obeyed $dA = n \,\Omega$ (see \cite{KN4}), so that
$d (n \omega^0 ) = ((k+1)/k) d A$ and
\beq
\alpha\, (n+ {\half}(k+1)) \omega^0 = \alpha\, {k+1\over k}
\left( A + {k \over 2} \omega^0 \right)
\label{met38}
\eeq
Fluctuations in the fields can be introduced at this stage and so, we have arrived at the
following conclusion
about the shift of $A$: {\it The implementation of
the symplectic transformations (\ref{met33}), which is itself rooted in the need to
allow for different choices of coordinates before reduction to
$\mathbb{CP}^k$, naturally leads to the combination
$A + {k \over 2} \omega^0 $ observed in the effective action}.

Now once again, we start with (\ref{met28}) and consider its geometric quantization.
The symplectic potential corresponding to $M$ is
\beq
{\cal A} = {i \over 2} ( Z_\alpha \, d{\bar Z}_\alpha -  {\bar Z}_\alpha \, d Z_\alpha )
\label{38a}
\eeq

Under canonical transformations (which preserve $M$),
${\A}$ transforms as $ {\A} \rightarrow
{\A} + d \, f$ (for some function $f$), thus behaving as a $U(1)$
gauge field. The wave functions are charged under this $U(1)$, transforming with a phase.
One must also consider covariant derivatives of the form
$\D_\alpha \Psi = (\del_{\alpha} -i  {\A}_{\alpha} ) \Psi$,
${\bar \D}_\alpha \Psi = (\bdel_{\alpha} -i  {\bar \A}_{\alpha} ) \Psi$
 in formulating the polarization condition.
The wave functions are thus sections of a holomorphic line bundle
on $\mathbb{C}^{k+1}$ with curvature $M$.

Explicitly, the covariant derivatives are
\beq
\D_\alpha = {\del \over \del Z_\alpha } - {\half} {\bar Z}_\alpha, \hskip .2in
{\bar \D}_\alpha = {\del \over \del {\bar Z}_\alpha } + {\half} {Z}_\alpha
\label{38b}
\eeq
The polarization condition 
$ \bar{\D}_\alpha \Psi = 0$ on the prequantum wave functions leads to the
coherent states (\ref{met30}), with the inner product defined by
the symplectic (Liouville) volume element for the phase space,
\beq
\la 1\vert 2\ra = \int \prod_\alpha dZ_\alpha d {\bar Z}_\alpha ~ 
e^{- {\bar Z}\cdot Z} ~ \Psi_1^* \, \Psi_2
\label{met38c}
\eeq
The prequantum operator corresponding to a function $f$ on the phase space is defined
in geometric quantization as \cite{geom, woit, nair-arm}
\beq
\P (f) = -i X \cdot \D + f 
\label{met38d}
\eeq
where $X$ is the vector field corresponding to $f$ defined by
$X^\mu \, M_{\mu \nu} = - \del_\nu f$.
We find easily that $X_{Z_\alpha} = - i\, ({\del/ \del {\bar Z_\alpha}})$,
$X_{{\bar Z}_\alpha} = i\, ({\del/ \del {Z_\alpha}})$.
The action of the corresponding prequantum operators are
\beqar
\P (Z) \Psi &=& ( - \D_{{\bar Z}_\alpha} + Z ) \Psi = e^{- {\half}{\bar Z}\cdot Z} ~ Z \, h(Z)\nonumber\\
\P ({\bar Z}) \Psi &=& (\D_{Z _\alpha} + {\bar Z}_\alpha ) \Psi =
e^{- {\half} {\bar Z}\cdot Z} ~ {\del \over \del Z_\alpha}  \, h(Z)
\label{met38e}
\eeqar
This is consistent with the assignment of $Z_\alpha$ as $a^\dagger_\alpha$ and
${\bar Z}_\alpha $ as $a_\alpha$. 

This is all standard and well known.
However, this picture of quantization is known to be incomplete. On a general symplectic
manifold, we can consider other polarizations, not necessarily the holomorphic one.
For example, on a phase space which is the cotangent bundle $T^*M$
of a real manifold
$M$, one can consider wave functions in the coordinate representation.
(This possibility applies to the present case as well, since we can consider
$\mathbb{R}^{ 2k + 2}$ as the cotangent bundle of $\mathbb{R}^{k+1}$.)
In such cases, because $\Psi^* \Psi$ depends only on half of the phase space coordinates, one has
to use a volume element on the subspace of such coordinates 
to define the inner product
for the wave functions.

The problem is that there is no such volume element defined by the given data
on the phase space.
The phase volume is naturally defined (in terms of powers of the symplectic
structure)
and can be used for the holomorphic polarization (for which
$\Psi^* \Psi$ depends on all phase space coordinates in general). 
But for real polarizations
the phase volume is not appropriate.
On the other hand, we would like to formulate quantization in a way which applies
to any choice of polarization, since physical results should be independent of polarization
(even though we may not have a real polarization for manifolds of interest).
One solution is to introduce ``half-forms" whose transformation property is such that
the product of two such forms transforms as the volume form of the submanifold
over which $\Psi^* \Psi$ is to be integrated.
We then consider the product of the line bundle (with curvature equal to the symplectic
two-form) and a bundle of half-forms, the wave functions being identified as sections of this
product bundle.
The transformation property of half-forms implies defining a square root of the Jacobian
of a symplectic diffeomorphism, so that at the level of linear transformation, we need to consider a double cover of the symplectic group, which is named the metaplectic group.

For the case of holomorphic polarization, which is our focus here, seemingly one can avoid 
using half-forms since the volume element for the full phase space can be used in the inner product.
However, the half-forms do add certain terms to the expressions for the operators; these additions are the ``metaplectic corrections". 
We want to argue that in the combination
$A + {k\over 2} \omega^0$, the
second term arises from such a correction.

The main point is that, generally, for all polarizations, the wave functions are of the form
\beq
\Psi \sim e^{- {\half} {\bar Z}\cdot Z}\, h(Z)~\sigma_{-1/2}(Z)
\label{met27a}
\eeq
where $\sigma_{-1/2} (z)$ indicates the appropriate section of the half-form.
For the case of holomorphic polarization, we do not need to know
an explicit form for $\sigma_{-1/2}(Z)$, only its transformation property is
important. In fact, we
may think of it
as a pure phase, which would not affect the inner product. However, the
vector fields corresponding to a function can have a nontrivial action on
$\sigma_{-1/2}(Z)$, and so the expression for the operator
 has  to be modified.
 With the half-form
$\sigma_{-1/2}$, this is given by \cite{{geom},{woit}, {nair-arm}}
\beq
\P (f) \, \Psi \, \sigma_{-1/2}  = \left[ \left( -i X\cdot \D + f \right) \Psi \right]\, \sigma_{-1/2}
- \Psi ~ ( i L_X \sigma_{-1/2} )
\label{met27b}
\eeq
where $L_X \sigma_{-1/2}$ is the Lie derivative of $\sigma_{-1/2}$ 
with respect to $X$.
Explicitly, if $X$ preserves the polarization,
we must have
\beq
[X, (\del / \del {\bar Z}_\alpha)] = C_\alpha^{~\beta} \,(\del/ \del {\bar Z}_\beta)
\label{met36aa}
\eeq
For such cases, one can show that
\beq
-i L_X \sigma_{-1/2} = - i X \cdot \del \sigma_{-1/2} - {i \over 2} \del \cdot X\, \sigma_{-1/2}
= -{i \over 2} \Tr\,C\, \sigma_{-1/2}
\label{met36b}
\eeq
where $\Tr\,C = C_\alpha ^\alpha$.
This shows that it is possible to view $\P (f)$ as acting just on $\Psi$
according to
\beq
\P (f) \, \Psi = \left[ \left( -i X\cdot \D + f \right)  -{i \over 2} \Tr\,C \right] \, \Psi
\label{met36c}
\eeq
and reabsorb $\sigma_{-1/2}$ and its conjugate into the measure of integration,
where they cancel out leaving just the phase volume defined by the symplectic form.
The extra term $-{i \over 2} \Tr\,C $ in $\P(f)$ is the metaplectic correction. The vector fields $X_{Z_\alpha},~X_{\bar{Z}_\alpha}$ commute with the polarization, $C_\alpha^\beta=0$ in (\ref{met36aa}) for these vectors, so the expressions for the quantum version of $Z_\alpha,~\bar{Z}_\beta$ are unchanged. However, for the vector field corresponding to $\bar{Z}_\alpha Z_\alpha$, (\ref{met36aa}) gives
 $C_\alpha^\beta =i\, \delta_\alpha^\beta$ and hence
\beq
\P ({\bar Z}_\alpha Z_\alpha ) \, \Psi
= e^{- {\half} {\bar Z}\cdot Z} \, \left( {Z_\alpha {\del \over \del Z_\alpha} }+ 
{\half} (k+1) \right) \, h (Z)
\label{met36d}
\eeq
This is equivalent to saying that the quantum operator corresponding to
${\bar Z}_\alpha Z_\alpha$ is $a^\dagger\cdot a + {\half} (k+1)$, which is the $Q$ we obtained previously in (\ref{met35}) in terms of the $Sp(k+1, \mathbb{R})$ algebra.
We see that the ``zero-point" charge can indeed be interpreted as the metaplectic
correction.

Here we have pursued the description of $\mathbb{CP}^k$ as the reduced phase space
obtained via symplectic
reduction from $\mathbb{C}^{k+1}$. This simplified the analysis
since the geometric quantization of
$\mathbb{C}^{k+1}$ is fairly straightforward and we could then use
those features
which descend to
$\mathbb{CP}^k$ to arrive at the combination $A + {k \over 2} \omega^0$.
But a direct geometric quantization of $\mathbb{CP}^k$ is also possible.
The identification of $\sigma_{-1/2}$ within such an approach
and the direct calculation of
the metaplectic correction
(or the $Mp^c$ correction \cite{rawn}) would be very interesting.

\vskip .2in
We thank Peter Woit for many useful comments, discussions and a careful 
reading of the manuscript.
This research was supported in part by the U.S.\ National Science
Foundation grants PHY-1417562, PHY-1519449
and by PSC-CUNY awards. 


\end{document}